\documentclass[conference]{IEEEtran}
\IEEEoverridecommandlockouts
\usepackage{cite}
\usepackage{amsmath,amssymb,amsfonts}
\usepackage{algorithmic}
\usepackage{graphicx}
\usepackage{textcomp}
\usepackage{geometry}
\usepackage{multirow}
\usepackage{pdflscape}
\usepackage{slashbox}
\usepackage{booktabs} 
\usepackage{authblk}
\usepackage{setspace}
\usepackage{subcaption}
\usepackage{wrapfig}
\usepackage{multirow}
\usepackage{multicol,caption}
\usepackage{afterpage}
\usepackage{wrapfig}
\newgeometry{paper=a4paper, left=0.550in, right=0.900in, bottom=1in, top=0.72in}
\afterpage{\aftergroup\restoregeometry}

\usepackage{graphicx}
\usepackage{titlesec}
\usepackage{multirow,tabularx} 
\usepackage{adjustbox} 
\usepackage{booktabs}
\usepackage{color, colortbl} 

\usepackage{xcolor}
\def\BibTeX{{\rm B\kern-.05em{\sc i\kern-.025em b}\kern-.08em
    T\kern-.1667em\lower.7ex\hbox{E}\kern-.125emX}}
\begin{document}

\title{Cultural-aware Machine Learning based Analysis of COVID-19 Vaccine Hesitancy}

\author[1]{Raed Alharbi}
\author[2]{Sylvia Chan-Olmsted}
\author[3]{Huan Chen}
\author[1*]{My T. Thai\thanks{*Corresponding author.}}
\affil[1]{Computer and Information Science and Engineering Department, University of Florida, Gainesville, FL, 32611, USA.}
\affil[]{\{r.alharbi, mythai\}@ufl.edu}

\affil[2]{Department of Media Production, Management, and Technology, University of Florida, Gainesville, FL, 32611, USA.}
\affil[2]{chanolmsted@jou.ufl.edu}
\affil[3]{Department of Advertising, University of Florida, Gainesville, FL, 32611, USA. }
\affil[3]{huanchen@jou.ufl.edu}

\renewcommand\Affilfont{\itshape\small}
\newcommand{\mt}[1]{\textcolor{red}{\# #1 (MT) \# }}   

\maketitle

\begin{abstract}
Understanding the COVID-19 vaccine hesitancy, such as who and why, is very crucial since a large-scale vaccine adoption remains as one of the most efficient methods of controlling the pandemic. Such an understanding also provides insights into designing successful vaccination campaigns for future pandemics. Unfortunately, there are many factors involving in deciding whether to take the vaccine, especially from the cultural point of view. To obtain these goals, we design a novel culture-aware machine learning (ML) model, based on our new data collection, for predicting vaccination willingness. We further analyze the most important features which contribute to the ML model's predictions using advanced AI explainers such as the Probabilistic Graphical Model (PGM) and Shapley Additive Explanations (SHAP). These analyses reveal the key factors that most likely impact the vaccine adoption decisions. Our findings show that Hispanic and African American are most likely impacted by cultural characteristics such as religions and ethnic affiliation, whereas the vaccine trust and approval influence the Asian communities the most. Our results also show that cultural characteristics, rumors, and political affiliation are associated with increased vaccine rejection.
\end{abstract}

\begin{IEEEkeywords}
COVID-19, Social Network
\end{IEEEkeywords}



\section{Introduction}


A better understanding of individuals and ethnic groups attitudes towards vaccination hesitancy is essential since it can provide insights into: (1) The distinct cultural behavior associated with vaccine-reluctant individuals and ethnic groups. (2) The impact of ethnic groups on individual members' decisions against vaccination, and (3) The characteristics shared by vaccine skeptics (when/why/how do they decide to accept/refuse vaccination). These insights, in turn, help us: (1) Design a model to predict the vaccine hesitancy for a given individual; and (2) develop efficient methodologies to vaccinate those who are reluctant and help in containing the current disease from spreading and combating future epidemics. Hence, it is very critical to have a full grasp of the reasons for vaccine hesitancy.

Unfortunately, vaccine hesitancy is a complicated and context-dependent phenomenon that is influenced by factors variable across time and location. Vaccine hesitancy can be associated with social-economic background (e.g., education, income, and occupation), perceptions of susceptibility (e.g., government, healthcare system, and doctors trust), and severity of disease (e.g., long-term side effects and health consequences). Furthermore, vaccination hesitation is not self-contained and can be influenced by a variety of variables such as the fast development and approval of COVID-19 vaccines, side effects shown in some individuals, and deaths of individuals after vaccination due to other complications. Some individuals use these factors to promote disinformation, adding to the long-standing distrust of public health institutions among various ethnic groups.

To tackle these challenges, we first focus on building a vaccine hesitancy prediction machine learning model called \textsc{VacAdopt}, with an emphasis on culture-awareness. In doing so, we conduct a systematic survey collecting $125$ possible factors underlying vaccine hesitancy. This new data set consists of comprehensive features ranging from vaccination health concerns to culture-related characteristics. We next focus on analyzing the important factors that impact an individual's decision on taking vaccination. This is equivalent to finding important features of an input that constitute to the \textsc{VacAdopt}'s prediction, which can be done using recent developments in explainable AI (XAI). 

XAI research has been emerged recently to provide explanations for a given input, in a form of important features, for any given black-box ML model in a model-agnostic fashion. Two notable explainers which this paper use are SHAP\cite{lundberg2017unified} and PGM \cite{vu2020pgm}. We perform an in-depth analysis using SHAP to identify key characteristics independently associated with COVID-19 vaccine acceptance or rejection. We next turn our attention to analyzing the composite features of vaccination willingness. To identify the correlation between important features, we adopt PGM explainer, which is developed for graph neural networks to work on our tabular dataset. Finally, we analyze the greater impact of particular features on specific ethnic groups, including Asian, African American, and Hispanic.


\begin{figure*}[ht]
    \centering
    \includegraphics[width=1.3\columnwidth]{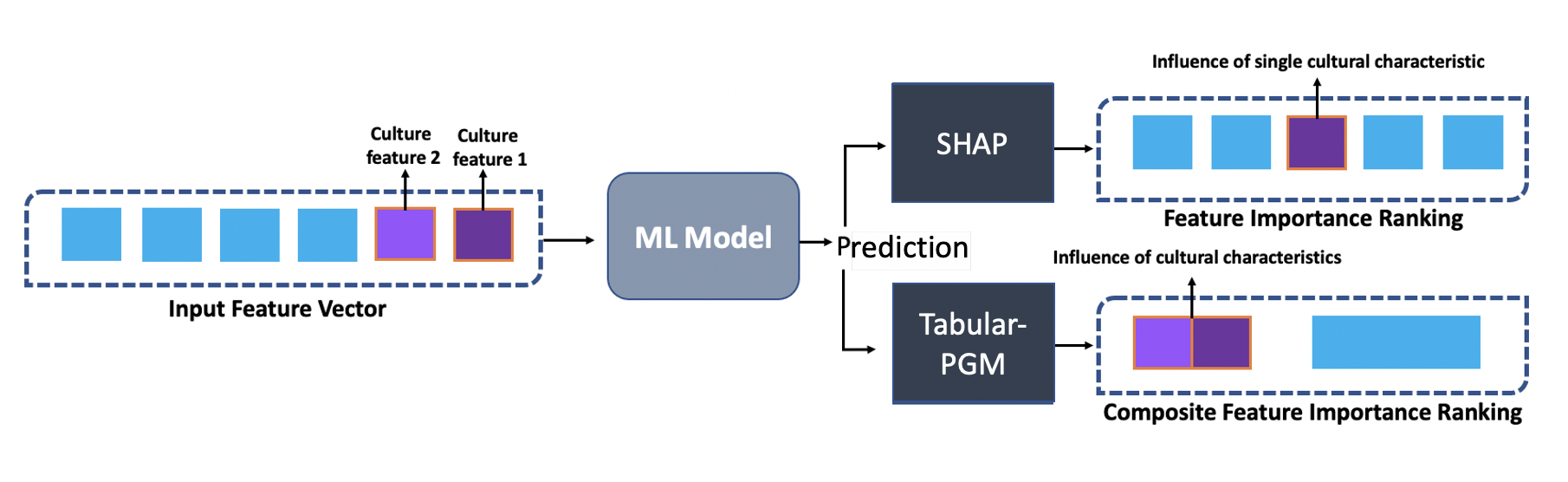}   
    \caption{\textsc{VacAdopt}: Cultural-awareness model architecture.} 
    \label{arch}
\end{figure*}



\textbf{Organization.} The remainder of the paper is structured as follows. Section \ref{related} presents the background and related works. The proposed framework details are introduced in section \ref{model_details}, while the experimental analysis is discussed in section \ref{exp_details}. Finally, section \ref{con} concludes the paper.

\section{Background and Related Work}
\label{related}

\textbf{Vaccination Willingness Models for Tabular Data.} The most frequent data format in real-world applications is tabular data, consisting of samples (rows) and a set of properties (columns). Tabular data is utilized across practical applications, including medical, banking research, and many other relational database-based applications. Due to their superior performance, conventional ML methods have remained dominant in tabular data applications over the past decade.

In particular, eXtreme Gradient Boosting (XGB) \cite{chen2016xgboost} is the state-of-the-art ML model for tabular data. XGB integrates numerous weak predictors, usually Decision Trees, to build an additive predictive model. Another technique is the decision tree \cite{safavian1991survey}, which presents feature-based split predictions in a flowchart-like tree. Other ML methods include k-Nearest Neighbors (KNN) \cite{peterson2009k}, which classify data based on its surrounding data points, and random forest (RF) \cite{breiman1999random}, an ensemble of classifiers built with randomization sources. 

Our work is related to exploring the above methods for COVID-19 vaccination willingness detection. Yet, there needs to be more characterizing comprehensive features ranging from vaccination health concerns to culture-related characteristics. Therefore, we first conduct a systematic survey collecting 125 possible factors underlying vaccine hesitancy. We next focus on developing a vaccine hesitancy prediction ML model, with a focus on cultural awareness.

\textbf{Explainable Machine Learning.}
Regardless of how widely ML models are adopted, they remain black boxes to users. Thus, ML explainers are employed to deduce (1) Why the selected models generated such predictions on a given input. (2) The model's behavior enables us to identify situations where the feature may significantly influence one ethnic group more than another. (3) The model's transparency increases model trust. ML explainable methods can be categorized into two types: post-hoc explainability and intrinsic explainability \cite{alharbi2021evaluating}. Post-hoc accomplishes interpretability using various statistical approaches to understand the model's behavior fully \cite{lundberg2017unified}. In contrast, intrinsic explainability is obtained by reducing the complexity of the model\cite{shu2019defend}.

To understand the reasoning behind models' predictions, we select two well-known explainers, SHAP \cite{lundberg2017unified} and PGM-Explainer \cite{vu2020pgm}. SHAP is designed for Conventional Neural Networks (CNNs), while PGM-Explainer is 
for Graph Neural Networks (GNNs).

In more details, SHAP is a game-theoretic technique to compute Shapley values to explain each feature's contribution to a prediction. The Shapley values of the model's conditional expectation function are approximated using additive feature attribution methods. And PGM-Explainer is an explanation approach that uses a graphical model to approximate the target prediction; PGM-Explainer demonstrates the effects of given features non-linearly to a prediction. However, due to its perturbation architecture, the current version of the PGM explanation is not directly usable on tabular data. Therefore, we propose a perturbation schema to use PGM-Explainer to handle our tabular dataset.

\section{The Proposed Framework - VacAdopt}
\label{model_details}

This section provides the implementation details of our proposed model, \textsc{VacAdopt} in Figure \ref{arch}, where the key components are: (i) A new benchmark that collects and evaluates many aspects of vaccination reluctance in a systematic manner, with an emphasis on culture-awareness features. (ii) A proof-of-concept ML model for predicting a user’s desire to receive COVID-
19 vaccination based on various baseline variables. (iii) And the adopted version of PGM (Tabular-PGM) to explain the influence of composite cultural features. 

This section starts with a description of the dataset, followed by the experimental setup implementation details, including encoding schema, hyperparameter optimization. Next, we discuss in depth the adopted version of PGM.

\subsection{Data preparation and collection.}
\label{dapre}
Given the nature of this paper in investigating acceptance or refusal of COVID-19 vaccination among America’s ethnic minorities, an online survey is the most suitable approach for collecting data. Prior to the primary data collection, a pretest is conducted using Amazon's MTurk to evaluate the scale items' reliability. MTurk is a robust crowdsourcing tool that allows academics to collect data more effectively while maintaining data quality \cite{kees2017analysis}. Once the scale's reliability has been determined, a national online survey was distributed via Qualtric (an American experience management company that allows users to create professional surveys and generate reports) in the Fall of 2021. 

Audience panels from Qualtrics provide researchers with access to a nationally representative sample that closely represents America's demographic makeup. According to recent US census data, Hispanics account for $18.1\%$ of America’s racial makeup, followed by African Americans at $13.4\%$, and Asians at $5.9\%$ \cite{kim2021current}. Thus, the ratio of three ethnic minority groups in the sample is maintained using a soft quota by Qualtrics to reflect America’s racial demography best. 

Information is collected include but not limit to general media consumption/use, special COVID and language media consumption factors, COVID-19 information seeking, COVID-19 information value, Trust in primary health information source/media, CDC, FDA, vaccine and vaccination perception. The final dataset's structure, characteristics, and statistics can be found in \cite{raed2022}.

\subsection{Experimental Setup.}
\label{Exse}

The ML model in Figure \ref{arch} is implemented to construct a predictor of COVID-19 vaccination willingness:  whether a person would accept or refuse the COVID-19 immunization (target variable). We utilize five well-known ML-based classification techniques: XGB \cite{chen2016xgboost}, decision tree \cite{safavian1991survey}, k-KNN \cite{peterson2009k}, RF \cite{breiman1999random}, Support Vector Machine (SVM) \cite{zhang2006svm}. For a fair comparison, we consider a deep learning method, TABNET \cite{arik2021tabnet}, that is designed for tabular data. 

\textbf{Encoding Schema.} 
Our structured dataset consists of many columns, including numerical and categorical features. Finding an appropriate encoding schema in which categorical features are converted to numbers is critical for developing an effective and successful machine learning model. In our experiment, we encode our categorical features by combining two well-known encoding techniques, 
one-hot \cite{berry1998factorial} and label encoding \cite{hancock2020survey}. Developing such a hybrid technique is not trivial, as one encoder can work better than the other according to specific features.

We first divide the non-numerical features in our prescribed dataset into (1) order-based and (2) unorder-based. The order-based refers to the features where the order of the values in the feature column is matter such as level of agreement/disagreement with the vaccine. In contrast, The unorder-based represents the values of the feature in an irrelevant order. For example, the gender feature column's should have equal encoding representation for the values on it (male and female). Hence, we use label encoding to encode the order-based features and the one-hot encoder to encode the unorder-based features.


\textbf{Hyperparameter Optimization.} In our experiments, we employ a random search technique \cite{bergstra2012random} with $30$ rounds of $5$-fold cross-validation to identify the optimal hyperparameters for \textsc{VacAdopt} by generating random combinations of the hyperparameters. To investigate whether increasing the size of the k-folds used in cross-validation during randomized search influences predicted accuracy, we retrained the selected models using $10$-fold and $15$-fold cross-validation. The predictive performance of the $5$-fold and both $10$, $15$-fold cross-validation models are only marginally different by around $1\%$. Thus, we conduct our final analyses on the average of the three $5$, $10$, and $15$ fold cross-validation models. The complete list of final hyperparameters for the six models is provided in \cite{raed2022}.



\begin{figure}[]
    \centering
    \includegraphics[width=0.9\columnwidth]{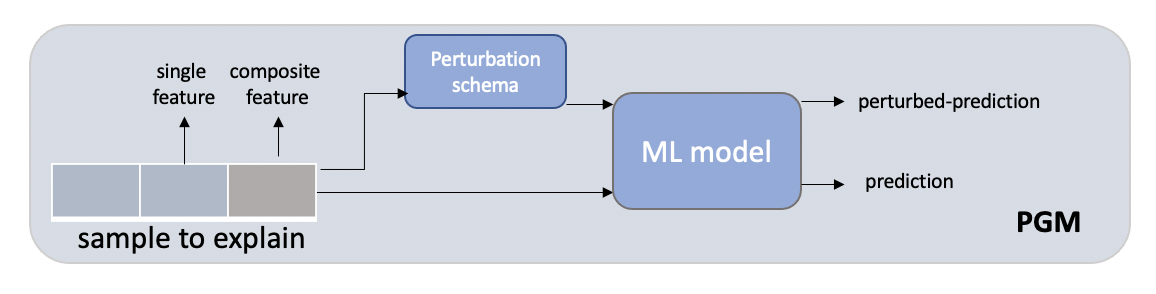} 
    \caption{Our Tabular-PGM explainer, adapting PGM-explainer\cite{vu2020pgm} to work on tabular datasets.}
    \label{tab-pgm}
\end{figure}

\begin{figure}[]
    \centering
    \includegraphics[width=0.9\columnwidth]{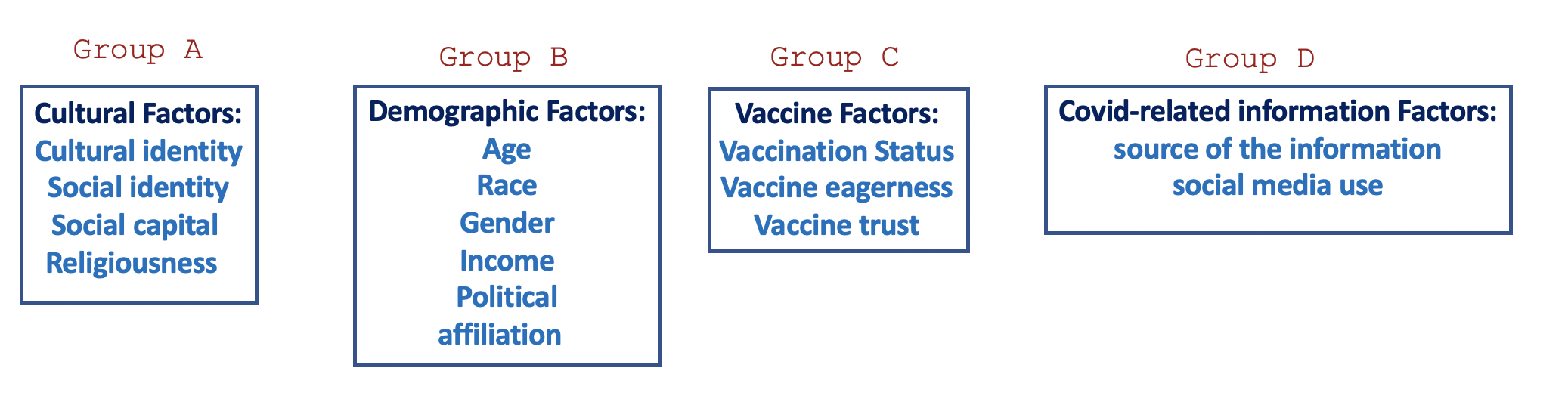}   
    \caption{The composite features are defined into distinct groups, including culture-based, demographic-based, vaccine-based, and COVID-19 related information factors. For simplicity, we have highlighted the key characteristics of each category.}
    \label{groups}
\end{figure}

\renewcommand{\arraystretch}{0.9}
\begin{table*}
\centering
\caption{Performance Comparison.}
\label{compare_fea_all}
\resizebox{\textwidth}{!}{%
\begin{tabular}{|lcccccccccccccccccccccccccccccccccr|}
\rowcolor{gray!60}
\multicolumn{1}{|c}{\textbf{Model \textbackslash{} Evaluation Metric}} &
  \textbf{Accuracy \%} &
  \textbf{F1-Score \%} &
  \textbf{Precision \%} &
  \textbf{Recall \%}

  \\
\multicolumn{1}{|l|}{\begin{tabular}[c]{@{}l@{}}\textbf{XGB Classifier - one hot encoding }    {}\end{tabular}} &

  \multicolumn{1}{c|}{$87.3$} &
  \multicolumn{1}{c|}{$92.2$} &
  \multicolumn{1}{c|}{$90$} &
  \multicolumn{1}{c|}{$94.3$}

    \\ \rowcolor{gray!30}
\multicolumn{1}{|l|}{\begin{tabular}[c]{@{}l@{}}\textbf{XGB Classifier - label encoding }    {}\end{tabular}} &

  \multicolumn{1}{c|}{$86.67$} &
  \multicolumn{1}{c|}{$91.7$} &
  \multicolumn{1}{c|}{$88.2$} &
  \multicolumn{1}{c|}{$95$}

  \\

\multicolumn{1}{|l|}{\begin{tabular}[c]{@{}l@{}}\textbf{Decision Tree Classifier}  {}\end{tabular}} &
  \multicolumn{1}{c|}{$82$} &
  \multicolumn{1}{c|}{$88.5$} &
  \multicolumn{1}{c|}{$90$} &
  \multicolumn{1}{c|}{$87.3$}
  
  \\ \rowcolor{gray!30}
\multicolumn{1}{|l|}{\begin{tabular}[c]{@{}l@{}}\textbf{KNN classifier}       {}\end{tabular}} &
  \multicolumn{1}{c|}{\textbf{$82.65$}} &
  \multicolumn{1}{c|}{\textbf{$89$}} &
  \multicolumn{1}{c|}{\textbf{$85$}} &
  \multicolumn{1}{c|}{\textbf{$94.4$}} 
  \\

\multicolumn{1}{|l|}{\begin{tabular}[c]{@{}l@{}}\textbf{Random Forest Classifier - one hot encoding}       {}\end{tabular}} &
  \multicolumn{1}{c|}{\textbf{$86$}} &
  \multicolumn{1}{c|}{\textbf{$92$}} &
  \multicolumn{1}{c|}{\textbf{$88$}} &
  \multicolumn{1}{c|}{\textbf{$96$}} 
   \\
    \rowcolor{gray!30}
   \multicolumn{1}{|l|}{\begin{tabular}[c]{@{}l@{}}\textbf{Random Forest Classifier - (label encoding + one hot encoding)}       {}\end{tabular}} &
  \multicolumn{1}{c|}{\textbf{88.8}} &
  \multicolumn{1}{c|}{\textbf{93.2}} &
  \multicolumn{1}{c|}{\textbf{$88.5$}} &
  \multicolumn{1}{c|}{\textbf{98}} 
   \\

\multicolumn{1}{|l|}{\begin{tabular}[c]{@{}l@{}}\textbf{SVM Classifier}       {}\end{tabular}} &
  \multicolumn{1}{c|}{\textbf{$87.13$}} &
  \multicolumn{1}{c|}{\textbf{$91$}} &
  \multicolumn{1}{c|}{\textbf{91}} &
  \multicolumn{1}{c|}{\textbf{$92$}} 
 
  \\
\rowcolor{gray!30}
\multicolumn{1}{|l|}{\begin{tabular}[c]{@{}l@{}}\textbf{TABNET}       {}\end{tabular}} &
  \multicolumn{1}{c|}{\textbf{$80$}} &
  \multicolumn{1}{c|}{\textbf{$88$}} &
  \multicolumn{1}{c|}{\textbf{$81.2$}} &
  \multicolumn{1}{c|}{\textbf{$96$}} 
 
  \\\hline
\end{tabular}}
\end{table*}

\subsection{Tabular-PGM explainer.}
\label{pgm_det}
Methodologically speaking, the perturbation part is a crucial step in XAI approaches \cite{lipton2018mythos} to obtain successful explainers. Perturbations provide a window into a black-box of DNN models by examining the input-output connection and determining which portion of the input is given special weight by the model (importance value). The PGM-explainer is a perturbation-based technique that concentrates on the correlation between various perturbed inputs and model outputs in a model-agnostic fashion. However, the perturbation in PGM-explainer is not applicable to be used on tabular dataset due to the fact that the PGM is designed for GNNs.

Therefore, to explain our tabular datasets, we adapted  PGM-Explainer to handle the tabular data, called Tabular-PGM explainer. Figure \ref{tab-pgm} shows our process to extract important feature values for a given sample, thereby giving an explanation in a form of which important features constitute a prediction of the model. The key difference between Tabular-PGM explainer and PGM-explainer lies in the perturbation scheme. In our Tabular-PGM explainer, we first set a variable $ v \in (0, 1)$ representing a probability that the features in each record are perturbed. For each record, we introduce a random variable $R$ indicating whether the features on $v$ are perturbed. The perturbed sample is then passed to the ML model to generate  \textbf{perturbed prediction}. The exact sample is also passed directly to the ML model to have the \textbf{prediction}. Finally, the important feature values are computed by observing the change in \textsc{VacAdopt}’s output given the perturbed examples using the probabilistic model described in PGM \cite{vu2020pgm}. The procedure is performed several times to minimize the effect of random permutations, and the feature values are averaged across five runs. 

Furthermore, since we are interested in knowing the influence of composite features on an individual's decision, 
we introduce a simple yet effective explanation method that adopts the PGM explainer to explain composite features. We merge multiple features into a single PGM node. In particular, we first define the composite features into distinct groups (Figure \ref{groups}). Then, the features inside the groups are treated as sensitive features. In other words, the perturbation schema in Figure \ref{tab-pgm} is implemented based on the presence of sensitive features in a group. For example, if a group contains two sensitive features (composite features), we would perturb the two features together to learn about their representation and effects.

\section{Experimental Analysis}
\label{exp_details}

Our goal is to first evaluate ML models in terms of predictive accuracy (section \ref{prePer}). Next, we conduct an in-depth analysis (section \ref{FeaImp}) using SHAP on the top predictive model to investigate (i) the key features behind an individual decision about COVID-19 and (ii) the correlation between certain features and vaccination acceptance or rejection. Our attention in section \ref{comfeim} is turned to focus on examining composite features using PGM-tabular to illustrate the influence of the related features, as the presence and effect of one feature could be derived from different factors. Section \ref{ethf} presents the details of identifying vaccination hesitancy characteristics of ethnic groups.


\subsection{Predictive performance.}
\label{prePer}

We begin by evaluating the performance of each detection model on the described dataset to determine its ability to predict user vaccination willingness using well-known evaluation metrics, including accuracy, precision, recall, and F1 score\cite{shu2019defend}. The evaluation and the comparison of the models are shown in Table \ref{compare_fea_all}. TABNET performs the worst in comparison to the other approaches, as demonstrated in Table \ref{compare_fea_all} (a higher number gives better results). This is because TABNET, a deep neural network, could not learn and represent characteristics that can aid in determining if a person was willing or reluctant to receive the vaccination. Deep neural networks, in general, do poorly with structured data. On the other hand, for the models designed to handle structured data (tabular dataset), we can see that the RF classifier (label+one hot encoding) outperforms XGB, the Decision Tree classifier, KNN, and SVM. For example, the RF classifier with hybrid encoding beats the second model in terms of the accuracy performance by about $1.7$.

\subsection{Features importance.}
\label{FeaImp}
\begin{figure*}
\begin{multicols}{2}

\begin{subfigure}{0.7\columnwidth}
  \includegraphics[width=\textwidth]{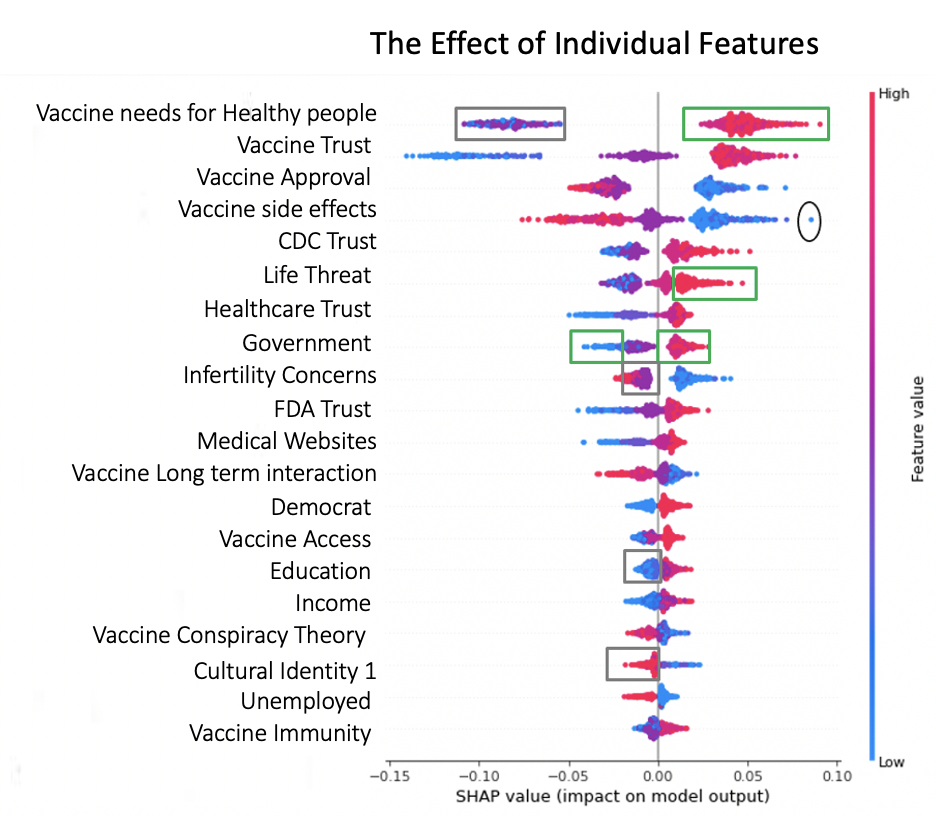}   %
    \caption{SHAP interpretation of the RF model} 
  \label{shap_rfc}
\end{subfigure}

\begin{subfigure}{0.7\columnwidth}
  \includegraphics[width=\textwidth]{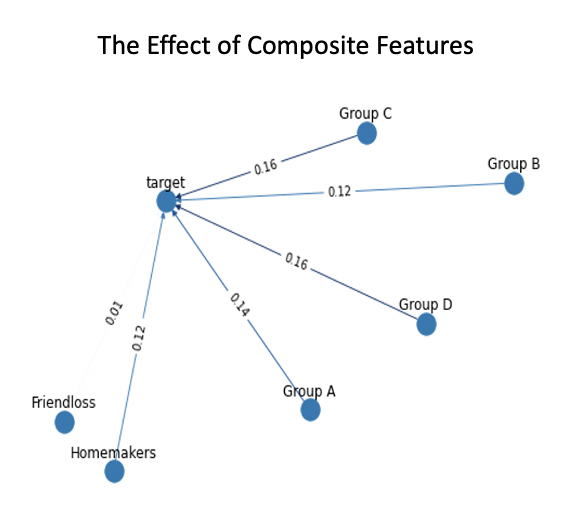}   %
  \caption{The adopted PGM interpretation of the RF model.} 

  \label{pgm_all}
\end{subfigure}
\end{multicols}

\caption{ \textbf{The top important features and composite features.}}
\label{global_ex}
\end{figure*}

Providing a trustworthy and faithful explanation of a model's behavior requires an effective and successful predictive model. Therefore, we select the model with the top performance (RF model) to be explained. Next, we examine the key features behind an individual decision about the COVID vaccine adoption using SHAP (Figure \ref{shap_rfc}). 

\begin{figure*}
\begin{multicols}{3}

\begin{subfigure}{0.885\columnwidth}
  \includegraphics[width=\textwidth]{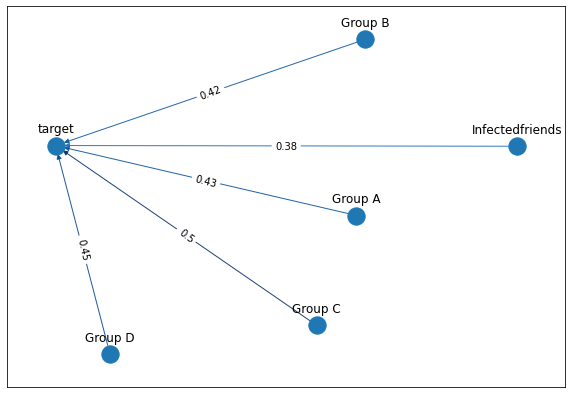}   %
  \caption{Asian Group}
  \vspace{0.8em}
  \label{asian}
\end{subfigure}

\begin{subfigure}{0.885\columnwidth}
  \includegraphics[width=\textwidth]{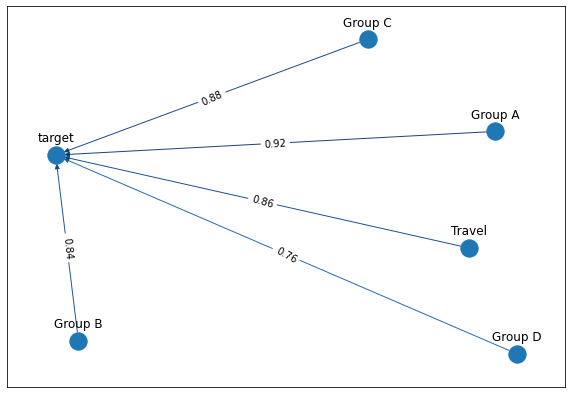}   %
  \caption{African American}
  \vspace{0.8em}
  \label{black_group}
\end{subfigure}

\begin{subfigure}{0.885\columnwidth}
  \includegraphics[width=\textwidth]{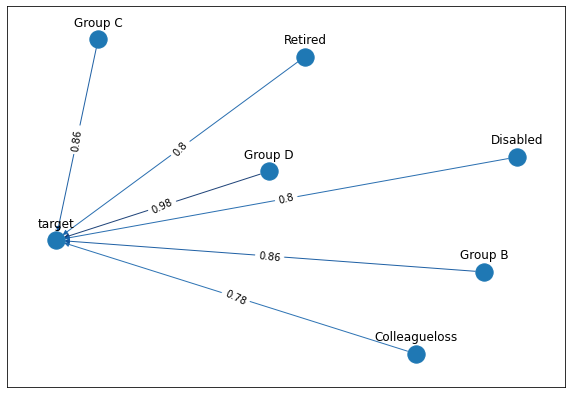}   %
  \caption{Hispanic Group}
  \vspace{0.8em}
  \label{hospanic_group}
\end{subfigure}
\end{multicols}
\caption{\textbf{The top important features/composite features for three ethnic groups, including Asian, African American, and Hispanic.}}
\label{pgm_group}
\end{figure*}

Figure \ref{shap_rfc} shows the top $20$ significant factors that impact a person's decision, where the features are listed according to their total contribution (sum of SHAP values) to the final prediction, with the top feature being the most important. Each point corresponds to a person in the study. The position of each point on the x-axis shows the impact that feature has on the classifier's prediction for a given individual, with colors ranging from red (high feature values) to blue (low feature values). On the x-axis, positive SHAP values indicate a willingness to receive COVID-19 vaccination, whereas negative SHAP values indicate vaccine reluctance. For example, the individual surrounded by a circle in Figure \ref{shap_rfc} illustrates that the vaccine side effect is not a big concern for the person (low feature value (blue)). Hence, the person decides to receive the vaccination (positive SHAP values).

We highlight here some of the most important $20$ features in Figure \ref{shap_rfc} (1-5 scale from strongly disagree to strongly agree): (1) Vaccine needs for healthy people: Do you think that COVID-19 vaccine is necessary for healthy people as well? (2) Vaccine Trust: Do you trust the COVID-19 vaccine? (3) Vaccine approval: Do not you trust the COVID-19 vaccine because of its current approval status? (4) Vaccine side effects: The possible side effects of the COVID-19 vaccine prevent me from receiving the vaccine. And (5) Conspiracy theory: an attempt to explain the COVID19 infection as the result of the actions of a minor, powerful group. Refer to \cite{raed2022} for more description of the features.

The vaccine-hesitant causes, using the SHAP explanations as shown in Figure \ref{shap_rfc}, can be divided into rational-based and irrational-based. The rational-based features are those that are debatable and understandable from the perspective of their health association. For example, the Food and Drug Administration (FDA) has authorized health organizations to use COVID-19 vaccines only in an emergency setting during the early phase of the pandemic. Unfortunately, such usage restrictions make individuals doubt the vaccine's effectiveness, thus refusing it (FDA trust in Figure \ref{shap_rfc}). The rapid development of the COVID-19 vaccine also raises key vaccine hesitancy issues such as vaccine trust, possible side effects, or long-term interaction that may pose a threat to people life, as illustrated in the RF interpretation in Figure \ref{shap_rfc}. If the right circumstances arise, rational-based cases can lead to vaccination. Several reasonable justifications are highlighted in green, shown in Figure \ref{shap_rfc}.


The irrational-based features, on the other hand, are those related to (1) cultural characteristics, (2) rumors, and (3) political affiliation. Although conspiracy theories  (considered as a specific type of rumors) are logically and scientifically implausible, they significantly influence vaccine adoption, as shown in Figure \ref{shap_rfc}. Another key feature is an individual's political background, or belief in governments. The uncertainty of certain groups might express itself in vaccination reluctance. More details can be seen in Figure \ref{shap_rfc}.

Looking closer, we notice that irrational-based features are strongly associated to the vaccine rejection. For instance, the majority of people who feel negative about their ethnicity (low feature value) refuse to obtain the vaccine. Some of the irrational-based features are highlighted in gray in Figure \ref{shap_rfc} to offer more insights into vaccine rejection. Overall, cultural characteristics, rumors, and political affiliation are associated with increased vaccine rejection.



\subsection{Composite features importance.}
\label{comfeim}
We now turn our attention to the composite factors instead of evaluating them independently. Grouping the characteristics can be used to illustrate the combined strength and influence of the related features, as the presence and effect of one main feature could be derived from different factors. For instance, cultural identity is developed and sustained via the exchange of common knowledge \cite{umana2004developing}. Hence,  a person's cultural identity cannot be formed only based on a single variable; rather, it can be formed based on numerous characteristics. Along this direction, we concentrate on the four main groups, as shown in Figure \ref{groups}. As long as the characteristics are relevant to one of the four major categories, they are assigned to that group.


Figure \ref{pgm_all} shows the PGM explanation for individual/composite features in the form of a graph network. The weights on the edges reflect the strength of the dependency connection between the target (receiving/refusing the vaccine predictions) and other nodes. The dependency connection shows how strongly that feature contributes to a prediction.) A high value indicates strong connections (dark blue), whereas a low value represents low connections (light blue).

As shown in Figure \ref{pgm_all}, the composite features in human health (represented by group C), such as COVID-related concerns, fear, vaccine trust, fertility problems, and side effects, remain the most prominent influences on vaccine adoption. However, the low vaccination rate has been motivated not only by significant healthcare concerns but also by the dissemination of various rumors and disinformation regarding the COVID-19 vaccine and its causes, as can be seen in Figure \ref{pgm_all} (group D has the same strength as group C). Differentiating and mitigating other rumors (group D) that take advantage of the government's long history of injustice can aid in the development of particular recommendations to boost vaccination rate acceptability.

Such a rise of social media misinformation (group D) as one of the primary sources or cultural factors (group A) as the second main composite feature for determining vaccination willingness decisions is a significant concern. Further investigations are needed to limit the misinformation and approach individuals with cultural-based attitudes as they impact by the people and environment.

\subsection{Ethnic groups features importance.}
\label{ethf}

Recognizing vaccination hesitancy characteristics of ethnic groups is crucial to understanding each group's concerns and need to accept the vaccine. Thus, for each ethnic group, we extract all data from the trained RF model that belong to that group and use Tabular-PGM explainer to analyze the key influences. As shown in Figure \ref{asian}, the vaccine factors (represented by group C) significantly influence vaccination adoption for Asian people. In contrast, the African American group is primarily influenced by cultural factors (represented by group A in Figure \ref{black_group}), such as cultural identity (e.g., group affiliation), social identity (e.g., socioeconomic status), and religiosity (e.g., following certain religions), whereas COVID-related information factors (represented by group D in Figure \ref{hospanic_group}), such as social media posts, are most important factors to the Hispanic group. We can also observe that the vaccination factors are the second most important composite feature for both the African American and Hispanic individuals (group C in Figures \ref{black_group} and \ref{hospanic_group}). This indicates that the trust in the vaccine is still a crucial concern for both groups but not as much as the cultural factors for African Americans and COVID-related information factors for Hispanics. 


These issues, when considered together, are well-established causes of mistrust aimed towards governments and the healthcare system, which impact vaccination uptake among various ethnic groups. The various key patterns among different ethnic groups provide insights into new strategies and methods for building trust, increasing collaboration, creating tools and resources to respond to the concerns and feedback from various ethnic groups. These efforts, in conjunction with scientifically backed information, can serve to boost COVID-19 vaccine acceptability.

\section{Conclusion} 
\label{con} 

This paper proposes a culture-aware machine learning (ML) model to predict vaccination willingness based on our new data collection. We adopt the advanced AI explainer (PGM) to work on a tabular dataset. We further analyze the most important features and composite features that contribute to the ML model’s predictions using advanced AI explainers. Our findings show that Hispanic and African Americans are most likely impacted by cultural characteristics, whereas the vaccine factors influence the Asian communities the most.

\section*{Acknowledgment}
This work was supported in part by the National Science Foundation Program under award No. 1939725

\bibliographystyle{IEEEtran}
\bibliography{Raedbibfile}
\end{document}